\documentclass[reprint, amsmath,amssymb,aps,pra]{revtex4-1}
\usepackage{subfigure}
\usepackage{graphicx}
\usepackage{dcolumn}
\usepackage{bm}
\usepackage{float}

\begin{document}
\title{Singlet state creation and Universal quantum computation in NMR using Genetic Algorithm}
\author{V. S. Manu } \email[manuvs@physics.iisc.ernet.in]{}
\author{Anil Kumar} \email[anilnmr@physics.iisc.ernet.in]{}
\affiliation{Centre for Quantum Information and Quantum Computing, Department of Physics and NMR Research Centre, Indian Institute of Science,  Bangalore}
\begin{abstract}
Experimental implementation of a quantum algorithm requires unitary operator decomposition. Here we treat the unitary operator decomposition as an optimization problem and use Genetic Algorithm, a global optimization method inspired by nature's evolutionary process for operator decomposition. As an application, we apply this to NMR Quantum Information Processing and find a probabilistic way of doing universal quantum computation using global hard pulses. We also demonstrate efficient creation of singlet state (as a special case of Bell state) directly from thermal equilibrium using an optimum sequence of pulses.
\end{abstract}
\maketitle
\section{Introduction} \label{sec:intro}

Quantum computation (QC) may possibly be the most remarkable proposal of practical application of quantum mechanics \citep{nielson}. Quantum information processor (QIP) is one which can control a large quantum system well enough to perform an arbitrary quantum algorithm. It holds out tremendous promise for efficiently solving some of the most difficult problems in computational science, such as integer factorization\citep{shor}, database search \citep{grover} and quantum simulation that are intractable on any present or future conventional computer\citep{feynman, aspuru_guzik, Benjamin, Jiangfengdu}.

Genetic algorithms (GA) are stochastic global search method based on the mechanics of natural biological evolution \citep{whitely}. It was first proposed by John Holland in 1975\citep{Holland}. GA operates on a population of solutions of a specific problem by encoding the solutions to a simple chromosome-like data structure, and applies recombination operators. At each generation, a new population is created by breeding individuals (selected according to their fitness value) together using operators borrowed from natural genetics. This process leads to the evolution of individuals and generate populations that are better suited to their environment. GAs are attractive in engineering design and applications because they are easy to use and are likely to find the globally best design or solution, which is superior to any other design or solution \citep{khaledrasheed}.

Long-lasting coherence and high fidelity controls in nuclear magnetic resonance (NMR) spectroscopy are ideal for quantum information processing. It made NMR QIP to perform first experimental implementations of basic quantum computation algorithms like Deutsch-Jozsa \citep{chuang_dj} algorithm, Grover's algorithm \citep{grover}, preparation of the three-qubit GHZ state \citep{laflamme_ghz}. Algorithms like shor's factorization algorithm \citep{shor}, dense coding\citep{bennet_dense}, quantum teleportation \citep{bennet_teleport} are demonstrated in NMR QIP and is one of the best candidate for testing basic principles of quantum mechanics \citep{jharana_nohiding, Mahesh_legget}.

Decomposing a unitary operator in terms of experimentally preferable operators is the main task in experimental implementation of a quantum algorithm. There are several proposals for doing this for NMR QIP, such as SMPS (Strongly Modulated Pulses) by Cory \emph{et al.}\citep{cory_smp}, GRAPE (Gradient Ascent Pulse Engineering) by Khaneja \emph{et al.}\citep{Khaneja_grape} and algorithmic approach by Ajoy \emph{et al.}\citep{ashok}. Here we investigate the use of Genetic algorithm for direct numerical optimization of pulse sequences and devise a probabilistic method for doing universal quantum computation by using hard pulses. We also investigate quantum state preparation using GA optimization. We demonstrate singlet state preparation (along with preparation of other three Bell states) directly from thermal equilibrium in two qubit homonulear NMR system by using global hard pulses and pulsed-field gradients \citep{price_pfg}. In Sec.\ref{sec:theory} of this paper, we describe the optimization procedure and in Sec.\ref{sec:exptl}, we  outline the experimental implementations.

\section{Genetic Algorithm for NMR Pulse Sequence Generation}  \label{sec:theory}
In liquid state NMR, the system Hamiltonian is composed of the interactions of spins with external magnetic field (chemical shifts) and coupling interactions among spins. Combining this with external RF pulses (with specific frequency, amplitude and phase) can simulate any preferred effective Hamiltonian \citep{ernstbook}. Hence unitary operator decomposition problem in NMR can be treated as an optimization problem which can give optimal values of pulse parameters and delay durations.  Here optimality is determined by a proper fitness function which depends on target Hamiltonian or target state.

	We have performed pulse sequence optimization using GA for quantum logic gates (operator optimization) and quantum state preparation  (state-to-state optimization)\citep{cory_smp}. State-to-state optimization converges faster than operator optimization (there can be many operators which can perform same state-to-state transfer). In the discussion given below, Single Qubit Rotation (SQR) pulses and CNOT gates are operator optimization whereas creation of pseudo pure state (PPS) and Bell states are state-to-state optimization.

\subsection{Representation Scheme:}  \label{sec:theory_n}

Representation scheme is the method used for encoding the solution of the problem to individuals in genetic evolution. Designing a good genetic representation that is expressive and evolvable is a hard problem in evolutionary computation \citep{whitely}. Defining proper representation scheme is the first step in genetic algorithm optimization.
	
In our representation scheme we have selected the gene as a combination of \emph{(i)} an array of pulses which will apply simultaneously on each channel with arbitrary amplitude ($\theta$) and phase ($\phi$) and \emph{(ii)} an arbitrary length delay $(d)$. It can be easily shown that repeated application of the above gene forms the most general pulse sequence in NMR. The individual, which represents a valid solution can be represented as a matrix of size $(n+1)\times2m$ and is shown in Eqn.\ref{eq:gene}. Here $m$ is the number of genes in each individual and $n$ is the number of channels or spins. So the problem is to find an optimized matrix (Eqn.\ref{eq:gene}) where optimality condition is posed by a fitness function (Sec. \ref{sec:ftns}).

\begin{equation}\label{eq:gene}
\begin{pmatrix}
\theta_{11}  & \phi_{11}  & . & . & \theta_{m1} & \phi_{m1}\\
\theta_{12}  & \phi_{12}  & . & . & \theta_{m2} &  \phi_{m2}\\
 . & . & . & . & . & . \\
 . & . & . & . & . & . \\
\theta_{1n}  & \phi_{1n}  & . & . & \theta_{mn} &  \phi_{mn}\\
d_1  & 0  & . &.  & d_1 & 0\\
\end{pmatrix}
\end{equation}

In the beginning of optimization, we select the number of genes ($m$) as an arbitrary number which is a function of the complexity of the problem to be solved. If the fidelity (fitness of the best individual in the population, calculated using fitness function Sec:\ref{sec:ftns}) crosses a cutoff value of (say $>99\%$), the optimization program tries to reduce the gene number by assigning zero value to gene parameters, and if not the program will rerun with more number of genes. In the optimization, we have used a population size of 100 individuals for 1000 generations. All the programs are written in $Matlab\textsuperscript{\textregistered}$ in combination with Matlab's optimization toolbox.

\subsection{Fitness Function:} \label{sec:ftns}

A fitness function is a particular type of objective function that prescribes the optimality of a solution or individual. In operator optimization, GA tries to reach a preferred target unitary operator ($U_{tar}$) from an initial random guess pulse sequence operator  ($U_{pul}$). Here we selected a fitness function which is proportional to the projection of the $U_{pul}$ into the target operator ($U_{tar}$).  It will give a maximum value of 1.0 when $U_{pul}$=$U_{tar}$ and the optimization has to run for maximizing the fitness function.

\begin{equation}  \label{eq:ftnso}
F_{pul}=Trace(U_{pul}\times U_{tar}^\dagger )
\end{equation}

In state-to-state optimization, the optimization program will run over different possibilities of $U_{pul}$ to prepare a preferred target state  $(\rho_{tar})$ from an initial state $(\rho_{in})$. The fitness function we have selected here is given by,

\begin{equation} \label{eq:ftnss}
F_{state}=Trace(U_{pul}\times \rho_{in} \times U_{pul}^{-1} \times \rho_{tar}^\dagger )
\end{equation}

Here also the optimization has to run for maximizing the fitness function.

\section{Two qubit homonuclear case:}  \label{sec:exptl}
\begin{figure}
\begin{center}
    \includegraphics[scale=0.7]{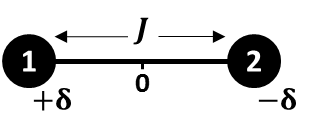}
    \caption{A twospin system with chemical shifts $\pm\delta$ and coupling  $J$}  \label{fig:twospn}
\end{center}
\end{figure}

Consider a two qubit NMR homonuclear system (Fig.\ref{fig:twospn}) with chemical shifts $\pm\delta$ and coupling  $J$.
Assuming weak coupling approximation $(\delta\gg J)$, the Hamiltonians can be written as\citep{ernstbook},

\begin{equation} \label{eq:csjh}
H=H_{cs}+H_J=2 \pi \delta (I_z^1-I_z^2)+2 \pi J (I_z^1I_z^2)
\end{equation}

For single qubit rotation (SQR) in the above system, one can use spin selective pulses (low power, long R.F pulses) which will excite a small spectral region around the spin to be selected \citep{mahesh_dj}.  On the other hand, by using the natural chemical shift difference between two spins, we show here how to implement SQR with global hard (non-selective) pulses. Later we extend this for performing two qubit homonuclear universal quantum computation using global hard pulses only.

\subsection{Operator Optimisation}
Operator optimization deals with  Pulse sequence generation for quantum logic gates. Here we show two essential unitary operators for universal quantum computation: Single Qubit Rotation (SQR) and two qubit Controlled-NOT gate.

\subsubsection{Single Qubit Rotations using non-selective pulses:} \label{sec:sqr}

For a two qubit homonuclear NMR system, we first consider the case $J=0$ (Eqn.\ref{eq:csh}).
 \begin{equation} \label{eq:csh}
H=H_{cs}=2 \pi \delta (I_z^1-I_z^2)
\end{equation}

 Evolution under such Hamiltonian can create a relative phase among spins. The relative phase accumulated is proportional to the chemical shift difference $(2\delta)$ and evolution time. Combining this relative phase with global rotation hard pulses, single qubit operations can be performed .

\begin{figure}
    \includegraphics[scale=1.2]{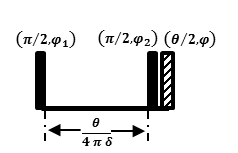}
    \caption{Pulse sequence for single qubit rotation.  First two filled pulses are $(\pi/2)$. Flip angle of the third pulse is $(\theta/2)$ with phase $\phi$. The brackets above each pulse contain the flip angle (first number) and the phase (second number).}
    \label{fig:sqrpp}
\end{figure}

The optimized pulse sequence for SQR is shown in Fig.\ref{fig:sqrpp}. In the beginning we had selected $m=3$, \emph{i.e.} three hard pulses and three delays. The optimised sequence has three hard pulses and single delay. The flip angle $(\theta)$ of the SQR pulse is determined by the delay and the flip angle of the third pulse, whereas the phase of SQR ($\phi$) and spin selection is determined by phases of all the three pulses (Tab.\ref{tab:sqrtab}).

\begin{table}
\begin{center}
\begin{tabular}{ccc}
\hline
\emph{Spin to be Excited} & $\phi_1$  & $\phi_2$ \\
\hline
1 & $(\phi-\pi/2)$ & $(\phi+\pi/2)$ \\
2 & $(\phi+\pi/2)$ & $(\phi-\pi/2)$ \\
\hline
\end{tabular}
\label{table1}
\caption{Spin selection and Phase $\phi$ of the SQR is controlled by the Phases $\phi_1$ and $\phi_2$ }\label{tab:sqrtab}
\end{center}
\end{table}

\begin{center}
\begin{figure}
 \subfigure[~]{{\includegraphics[width =0.75\linewidth,height =0.29\linewidth]{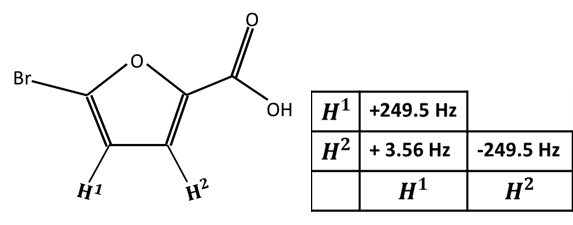}} \label{fig:bfra}}
 \subfigure[~]{{\includegraphics[width =0.7\linewidth,height =0.22\linewidth]{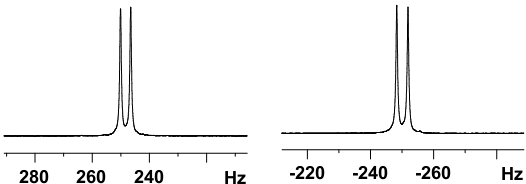}}  \label{fig:eqbmspectrum} }
 \subfigure[~]{{\includegraphics[width =0.7\linewidth,height =0.22\linewidth]{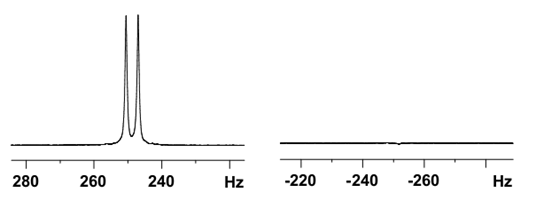}} \label{fig:sqr1} }
 \subfigure[~]{{\includegraphics[width =0.7\linewidth,height =0.22\linewidth]{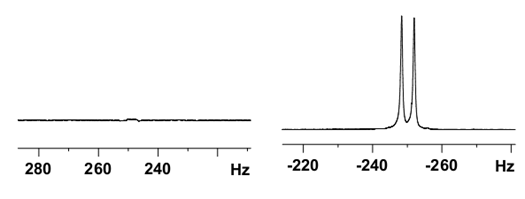}} \label{fig:sqr2}}
\caption{ (a). Chemical structure of 5-Bromofuroic acid. Diagonal elements in the table contains the chemical shifts of protons at 500 MHz and the   nondiagonal element represents J coupling (sample dissolved in $C_6D_6$). (b). Equilibrium Spectrum (c). $(\pi/2)_y$ SQR pulse on spin-1      (d). $(\pi/2)_y$ SQR pulse on spin-2. The average experimental fidelity (calculated using standard definition \citep{cory_smp}; spectral intensities were compared with equilibrium spectrum)  for the SQR pulse is obtained as 99.9\%.}
\end{figure}
\end{center}

Experimental verification of SQR pulse sequence in 5-Bromofuroic acid (Fig.\ref{fig:bfra}) (in $C_6D_6$) system is shown in Fig.\ref{fig:sqr1} and \ref{fig:sqr2}. The total length of the pulse sequence for $(\pi/2)$ SQR pulse is less than $500 \mu s$ where as conventional method (using a selective soft pulse) would need a $2ms$ shaped pulse. This shortening in time can lead to a significant improvement in a quantum circuits.

\begin{center}
\begin{figure}
  \includegraphics[width=0.7\linewidth]{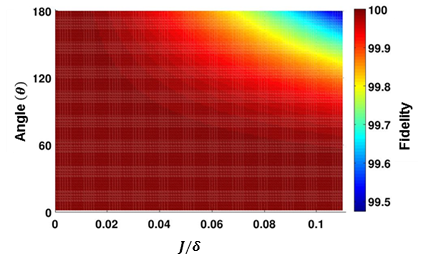}
  \caption{Matlab simulation study of Fidelity variation with different values $(J/\delta)$ and flip angle $(\theta)$. }\label{fig:sqrfv}
\end{figure}
\end{center}

The above analysis also holds for $J\neq0$, so long as $\gamma B_1\gg \delta, J$; except that introduction of $J$ coupling dephases the final state and results in a fidelity loss. The fidelity of the pulse sequence (fitness function Eqn.\ref{eq:ftnso}) is studied using Matlab simulation (Fig.\ref{fig:sqrfv}) and is $>99.8\%$ for $J/\delta < 0.1$ and $\theta < \pi/2$. ($J/\delta < 0.1$ is the limit for weakly coupled spins and in this paper we are dealing with only weakly coupled spins). For $ \pi/2 <\theta  < \pi$, the theoretical fidelity can reach upto 99.5\% (Fig.\ref{fig:sqrfv}).

\subsubsection{Controlled NOT gate (C-NOT)} \label{sec:cnot}

\begin{figure} [h]
\subfigure[~]{\includegraphics[width=0.85\linewidth]{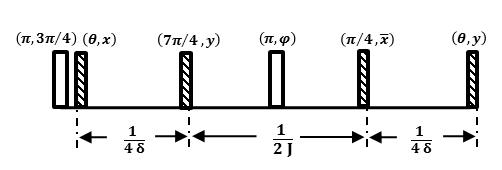} \label{fig:cnotpp}}
\subfigure[~]{\includegraphics[width=0.8\linewidth]{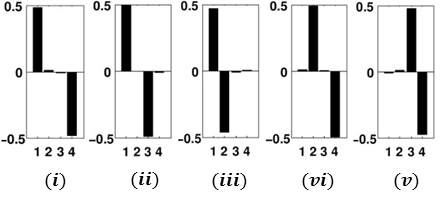} \label{fig:cnotspectrum}}
\subfigure[~]{\includegraphics[width=0.7\linewidth]{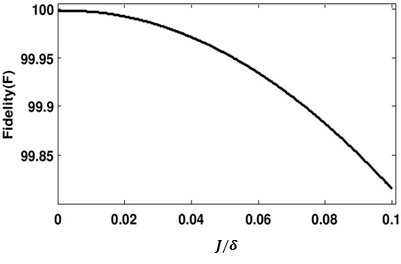}\label{fig:cnotfv}}
\caption{Pulse sequence for Controlled-NOT gate. The brackets above each pulse contain the flip angle (first number) and the phase (second number). (b) Diagonal element tomography of (i). Equilibrium state (ii). after applying C-NOT(1,2) (iii). C-NOT(2,1), (iv). C-NOT ($\overline{1}$,2) and (v). C-NOT ($\overline{2}$,1). Here \{1, 2, 3, 4\} corresponds to \{$|00\rangle$, $|01\rangle$, $|10\rangle$, $|11\rangle$\}. An average experimental fidelity of 99.9\% is observed. (c) The fidelity $(F) ~versus~ (J/\delta)$ plot of C-NOT gate.}
\end{figure}

The Controlled-NOT gate is an essential component in the construction of a quantum computer. Any quantum circuit can be simulated to an arbitrary degree of precision using a combination of C-NOT gates and single qubit rotations \citep{nielson}. The optimised pulse sequence for Controlled-NOT is shown in Fig.\ref{fig:cnotpp} (obtained using the Hamiltonian of Eqn. \ref{eq:csjh}). All the four C-NOT gates can be obtained by tuning $(\theta,\phi)$ as shown in Tab.\ref{tab:cnottab}. The pulse sequence is identical for all the four C-NOT gates except the angles $\theta$ and $\phi$.

\begin{table}
\begin{center}
\begin{tabular}{|c|c|c|}
  \hline
  \emph{Gate} & $\theta$ & $\phi$ \\
  \hline
  $CNOT(1,2)$ & $\pi/4$ & $\pi/2$ \\
  $CNOT(\overline{1},2)$ & $\pi/4$ & 0 \\
  $CNOT(2,1)$ & $3\pi/4$ & 0 \\
  $CNOT(\overline{2},1)$ & $3\pi/4$ & $\pi/2$ \\
  \hline
\end{tabular}
  \caption{($\theta$,$\phi$) values for all four CNOT gates} \label{tab:cnottab}
\end{center}
\end{table}

The experimental implementation of various C-NOT gates in 5-Bromofuroic acid is shown in Fig.\ref{fig:cnotspectrum}. An average experimental fidelity (calculated using \citep{cory_smp} with only diagonal element of the density matrix) of 99.9\% is achieved.

The theoretical fidelity of the operator (using Eqn.\ref{eq:ftnso}) is however dependent on the ratio $(J/\delta)$.  Matlab simulation of fidelity with $(J/\delta)$ for C-NOT gate pulse sequence is shown in Fig.\ref{fig:cnotfv}, and is $>99.99\%$ for $(J/\delta) = 0.01$ and $>99.84\%$ for $(J/\delta) <0.1$. This means that even if one needs 10 C-NOT gates in a quantum circuit, the fidelity is $>$99\%.

\subsection{State to State Optimization}
State to state optimization deals with pulse sequence generation for quantum state preparation. Here we added gradient pulses to the optimization procedure, which enables to perform non-unitary transformations and show two important quantum state preparations: Pseudo Pure state creation and Bell state creation directly from thermal equilibrium (mixed) state using global hard pulses.

\subsubsection{PPS creation}
Quantum information processing by NMR spectroscopy uses pseudo-pure states to mimic the evolution and observations on pure states \citep{cory_pps}. There are several methods for creating pseudo pure states (PPS) from thermal equilibrium \citep{cory_pps, gershenfeld_pps, Mahesh_sallt}.  Here we closely follow the Spatial Averaging Method (SAM) of Cory \emph{et al}. However, the SAM uses spin selective pulses, which in homonuclear system becomes soft long pulses. Here we obtain a novel pulse sequence using only non-selective (hard) pulses for homonuclear two qubit system (Eqn.\ref{eq:csjh}). The optimization problem here is a state to state optimization with thermal equilibrium state $\Delta\rho_{eq}=I_z^1+I_z^2$ as the initial state and $\Delta\rho_{00}=I_z^1+I_z^2+2I_z^1I_z^2$ as the final state.

\begin{figure}[!]
\subfigure[~]{{\includegraphics[width=0.9\linewidth]{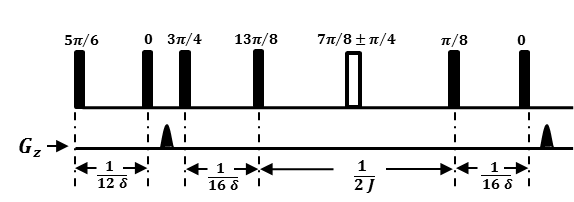}} \label{fig:ppspp}}
\subfigure[~]{{\includegraphics[width =0.76\linewidth,height =0.30\linewidth]{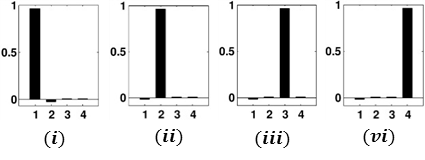}}\label{fig:ppsspectrum}}
\subfigure[~]{{\includegraphics[width=0.7\linewidth]{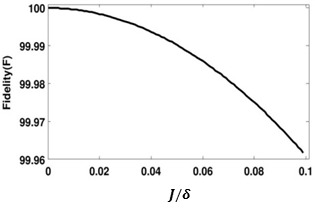}}\label{fig:ppsfv}}

\caption{(a)Pulse sequence for PPS creation. All filled pulses are  $(\pi/2)$  and nonfilled are $\pi$ with phases written above it. $\pm$ in the phase of $\pi$ pulse determines the pps to be created ($|00\rangle$ or $|11\rangle$). (b) Tomography of diagonal elements after preparing (i) $|00\rangle$ pps, (ii) $|01\rangle$ pps, (iii) $|10\rangle$ and (iv) $|11\rangle$ pps. Here \{1, 2, 3, 4\} corresponds to \{$|00\rangle$, $|01\rangle$, $|10\rangle$, $|11\rangle$\}. An average experimental fidelity of 99.5\% was obtained. (c) The fidelity $(F) - (J/\delta)$ plot of PPS generation pulse sequence (Fig.\ref{fig:ppspp}).}
\end{figure}

For easier optimization (and experimental implementation) we have fixed all the pulses to be $(\pi/2)$ and optimized only the pulse phases. The sequence consists of six $(\pi/2)$ pulses and one $(\pi)$ pulse for refocusing chemical shift (Fig.\ref{fig:ppspp}). The phase of the $(\pi)$  pulse can be controlled to achieve either $|00\rangle$ pps or $|11\rangle$  pps. The other PPS are obtained by using a combination sequence of PPS and SQR $\pi$ pulse. The experimental results are shown Fig.\ref{fig:ppsspectrum}. An average experimental fidelity (calculated using \citep{cory_smp} with only diagonal element of the density matrix) of 99.5\% is obtained for various PPS.

The theoretical fidelity (using Eqn.\ref{eq:ftnss}) of PPS preparation pulse sequence is also dependent dependent on the ratio $J/\delta$ and studied using Matlab simulation (Fig.\ref{fig:ppsfv}). The theoritical fidelity of the pulse sequence is $>99.9\%$ for $J/\delta < 0.1$.

\subsubsection{Creation of Bell states directly from equilibrium state}

\begin{figure*}
\subfigure[~]{{\includegraphics[width=0.8\linewidth]{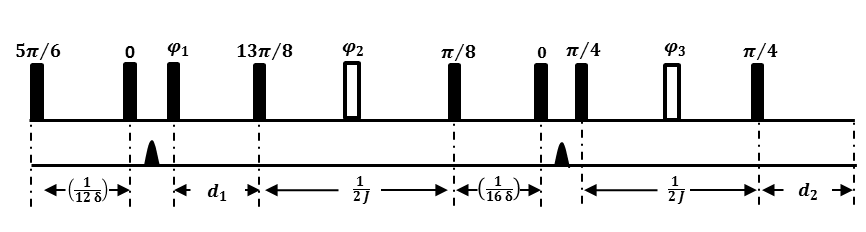}} \label{fig:singletpp}}
\subfigure[~]{{\includegraphics[width =0.4\linewidth]{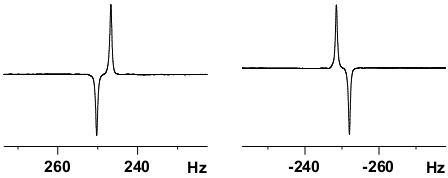}} \label{fig:singletexptspe}}
~~~~~~~~~~~~~~~~~~~~
\subfigure[~]{{\includegraphics[width =0.2\linewidth]{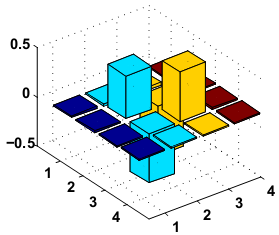}} \label{fig:singletexpttomo}}
\caption{(a). Pulse sequence for creating Bell states directly form thermal equilibrium state. All filled pulses are $(\pi/2)$ and nonfilled are $\pi$. The phase of each pulse is written above. The values of `$\theta$'s and `$d$'s are shown in Tab.\ref{tab:singlettab}, (b).Observing singlet state after applying $U(=e^{-i (\pi/2)I_x^{1,2}} \times  e^{-i (\pi/4)(I_z^1-I_z^2)})$, (c). Density matrix tomography of created singlet state.}
\end{figure*}

Bell states are maximally entangled two qubit states (also known as Einstein-Podolsky-Rosen states) \citep{epr}. They play a crucial role in several applications of quantum computation and quantum information theory. They have been used for teleportation, dense coding and entanglement swapping \citep{bennet_teleport, bennet_dense, pan_entswap}. Creation of Bell state is more demanding in quantum computation and conventionally (in NMR) it requires product state pps creation + Hadamard gate + C-NOT gate \citep{jharana_bell}. Here we integrated all these in a single pulse sequence and optimize with GA. Again we have kept all pulse amplitude to be $(\pi/2)$ and optimized for pulse phases and delay durations. The optimized pulse sequence (Fig.\ref{fig:singletpp}) has only ten non-selective pulses. The final Bell state can be decided by controlling the phase of the pulses and delay durations according to Tab.\ref{tab:singlettab}.

~

\begin{table}[H]
\begin{center}
\begin{tabular}{|c|c|c|c|c|c|}
  \hline
   \emph{Final Bell State} & $\phi_1$ & $\phi_2$ & $\phi_3$  & $d_1$ & $d_2$ \\
   \hline
   $|\psi^+\rangle=\frac{1}{\sqrt{2}}(|00\rangle+|11\rangle)$ & $3\pi/4$ & $9\pi/8$ & $3\pi/4$ & $1/16 \delta$ & $0$ \\
   $|\psi^-\rangle=\frac{1}{\sqrt{2}}(|00\rangle-|11\rangle)$ & $3\pi/4$ & $9\pi/8$ & $\pi/4$ & $1/16 \delta$ & $0$ \\
   $|\phi^+\rangle=\frac{1}{\sqrt{2}}(|01\rangle+|10\rangle)$ & $0$ & $5\pi/8$ & $3\pi/4$ & $9/48 \delta$ & $9/8 \delta$ \\
   $|\phi^-\rangle=\frac{1}{\sqrt{2}}(|01\rangle-|10\rangle)$ & $0$ & $5\pi/8$ & $\pi/4$ & $9/48 \delta$ & $9/8 \delta$ \\
  \hline
\end{tabular}
  \caption{The values of $\phi$ and $d$ in the pulse sequence shown in Fig.\ref{fig:singletpp}} \label{tab:singlettab}
\end{center}
\end{table}

The experimental implementation is performed in 5-Bromofuroic acid, for the Bell state $|\phi^-\rangle=\frac{1}{\sqrt{2}}(|01\rangle-|10\rangle)$, also known as the singlet state and is a long lived state for the Hamiltonian  $H=I_1.I_2$ \citep{levitt_lls,levitt_lls_2}.

\begin{equation}
\rho_{\phi^-}=(0.25 I- I_x^1I_x^2 - I_y^1I_y^2- I_z^1I_z^2)
\end{equation}

The experimental results are shown in Fig.\ref{fig:singletexptspe} and \ref{fig:singletexpttomo}. Since $\rho_{\phi^-}$ is not directly observable, we convert the created singlet state into observable single quantum coherence by applying $U=e^{-i (\pi/2)I_x^{1,2}} \times  e^{-i (\pi/4)(I_z^1-I_z^2)}$. An experimental fidelity (using the standard definition in \citep{cory_smp}) more than of 99.5\% is achieved. This is the shortest known pulse sequence for creating pure singlet state in a two qubit homonuclear NMR system.

\begin{figure}
  \includegraphics[width=0.7\linewidth]{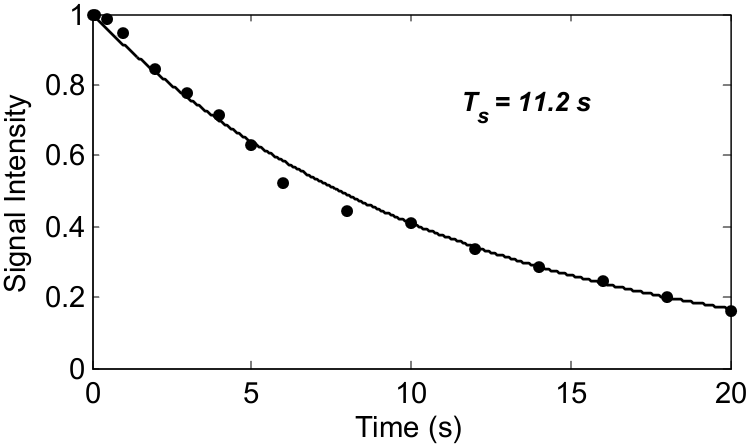}\\
  \caption{Anti-phase signal decay as a function of interval and fits to a single exponential decay. The initial amplitude of singlet state is normalized to one. Observed Singlet state life time is $T_s= 11.2~Sec$. The system has a $T_1=8.7 ~Sec$ and $T_2=3.8~Sec$.}\label{fig:singletlt}
\end{figure}

The singlet state life time ($T_s$) has been measured by applying Waltz-16 \citep{waltz16} spin lock sequence for a variable time period $(0-20~Sec)$ and is obtained as $11.2~Sec$ (Fig.\ref{fig:singletlt}) longer than $T_1=8.7~Sec$ and $T_2=3.8~Sec$.

Pulse sequence generation using GA for higher qubit systems is in progress. SQR and C-NOT gates using hard pulses (Sec. \ref{sec:sqr} and \ref{sec:cnot}) is valid for higher spin systems with homonuclear spin pairs (for example $^1H-^1H-^{19}F-^{19}F$ system in 2,3-Difluro-6-nitrophenol\citep{mahesh_dj}).

\section{Conclusion}

In conclusion, we have used the global optimization power of Genetic algorithm for \emph{(i)} efficiently implementing SQR and C-NOT gates and \emph{(ii)} creating PPS in homonuclear two qubit system using only hard pulses, and hence showed a method for doing universal quantum computation in such systems. We have also demonstrated the creation of Bell states directly from thermal equilibrium state. The pulse sequence for Bell state creation is the shortest known sequence. Unlike GRAPE \citep{Khaneja_grape} or SMPS \citep{cory_smp}, all the pulse sequences discussed here are valid for any weakly coupled homonuclear spin pairs.

~\\~\\
\bibliographystyle{unsrtnat}
\bibliography{ss_uqc}
\end{document}